\documentclass{ws-procs9x6}
\usepackage{psfrag,epsfig,mathrsfs}

\begin{document}

\newcommand*{\etal}{\emph{et al.}}
\newcommand*{\ddd}{\mathrm d^{3}}
\newcommand*{\dd}{\mathrm d}
\newcommand*{\ddk}[1]{\frac{\ddd k}{(2\pi)^3\, 2\omega_{#1}}}
\newcommand*{\vvec}[1]{{\bf #1}}
\newcommand*{\halb}{\frac{1}{2}}
\newcommand*{\del}{\partial}
\newcommand*{\Gl}[1]{(\ref{#1})}
\newcommand{\DSGl}{\textnormal{Dyson}-\textnormal{Schwinger} equation}

\title{Nonequilibrium Dynamics of the $O(N)$ linear sigma model in the
  Hartree approximation}

\author{S.~Michalski}

\address{Institut f\"ur Physik \\ Theoretische Physik III \\
  Otto-Hahn-Str. 4 \\
  D--44221 Dortmund, Germany \\
  E-mail: stefan.michalski@uni-dortmund.de}


\maketitle

\abstracts{
We investigate the out-of-equilibrium evolution of a classical background 
field and its quantum fluctuations in the scalar $O(N)$ model with
spontaneous symmetry breaking
\cite{Baacke:2001neqon}. We consider the 2-loop 2PI effective action
in the \textnormal{Hartree} approximation, i.e. including bubble resummation
but without non-local contributions to the \DSGl. 
We concentrate on the (nonequilibrium) phase structure of 
the model and observe a first-order transition between a
spontaneously broken and a symmetric phase at low and high energy
densities, respectively. So typical structures expected in thermal
equilibrium are encountered in nonequilibrium dynamics even at early 
times before thermalization. }


\section{The model}
\subsection{Applications}
Scalar models have a wide range of applications in quantum field
theory. Normally they are parts of more complex models like e.g.
the Standard Model or Grand Unified Theories but they often
serve as toy models for a simplified description of complex phenomena
such as inflationary cosmology or meson interactions in relativistic
heavy ion collisions.

\subsection{Nonequilibrium 2PI effective potential}
We consider the $O(N)$ model with spontaneous symmetry breaking 
whose classical action is
\begin{equation}
  \label{eq:action}
  \mathcal{S}[\vec\Phi] = \int \dd^4x\ \mathscr{L}[\vec{\Phi}]
   = \int \dd^4x\ \biggl\{
   \halb\ \del_\mu \vec{\Phi} \cdot \del^\mu \vec{\Phi} 
   - \frac{\lambda}{4} \left(\vec{\Phi}^2-v^2\right)^2 
   \biggr\}\ .
\end{equation}

Following Refs.~\cite{Baacke:2001neqon,Nemoto:1999onft}
we can compute the 2PI effective action \cite{Cornwall:2PI} 
in the Hartree approximation.
Furthermore, we diagonalize the Green function by an $O(N)$-symmetric ansatz
and by restricting the background field to one direction. Since the Green 
function is local, it can be described by two (time-dependent) mass parameters 
$\mathcal{M}^2_{\sigma,\pi}$.
For nonequilibrium purposes it can be factorized into 
mode functions
\begin{equation}
  \label{eq:Green-Funktion}
    G_{\sigma,\pi}(\vvec{x},t_>;\vvec{x}',t_<) = 
  \int \ddk{\sigma,\pi}\ f_{\sigma,\pi}(k,t_>)\, \bar{f}_{\sigma,\pi}(k,t_<)\
    e^{i \vvec{k} \cdot (\vvec{x}-\vvec{x}')}
    \ ,  
\end{equation}
where $\omega_{\sigma,\pi} = \sqrt{\vvec{k}^2+\mathcal{M}_{\sigma,\pi}^2(0)}$.
One constructs an expression for the total (conserved) energy density
of the system in the Hartree approximation
\begin{eqnarray}
  \nonumber
  \mathcal{E} &= & \halb \dot{\phi}^2 + \halb \mathcal{M}_\sigma^2 \phi^2
  - \frac{\lambda}{2} \phi^4 - \frac{v^2}{2 (N+2)} \biggl[
    \mathcal{M}_\sigma^2 + (N-1) \mathcal{M}_\pi^2 \biggr] \\
    \label{eq:EnergiedichteAux}
    && -\frac{1}{8\lambda (N+2)} \biggl[ (N+1) \mathcal{M}_\sigma^4 
      +3(N-1) \mathcal{M}_\pi^4 - 2(N-1) \mathcal{M}_\sigma^2\ \mathcal{M}_\pi^2\\
    \nonumber
    &&  \qquad\qquad + 2N\lambda^2 v^4 \biggr] 
    + \mathcal{E}_{\mathrm{fl}}^\sigma(t)
    + (N-1)\,\mathcal{E}_{\mathrm{fl}}^\pi(t)\ ,
\end{eqnarray}
where the fluctuation energy densities $\mathcal{E}_{\mathrm{fl}}^{\sigma,\pi}$ 
are the nonequilibrium analogs
of the one-loop ``log det'' terms expressed by mode functions $f(k,t)$
\begin{equation}
  \label{eq:Efluct}
  \mathcal{E}_{\mathrm{fl}}^*(t) =
  \frac{\hbar}{2} 
  \int\ddk{*}\ \biggl[ |\dot{f}_*(k,t)|^2 + 
  (\vvec{k}^2 + \mathcal{M}_{*}^2)\ |f_*(k,t)|^2
    \biggr],\  *=\sigma,\pi\ .
\end{equation}

\subsection{Equations of motion}
The equations of motion follow from the conservation of the 
energy~\Gl{eq:EnergiedichteAux}. The background field obeys 
\begin{equation}
  \label{eq:bewgl phi_i}
  \ddot{\phi} + \bigl[ \mathcal{M}_\sigma^2(t) - 2\lambda\ \phi^2(t) 
  \bigr] \phi(t) = 0\  ,
\end{equation}
the mass parameters are solutions of the gap equations
  \begin{eqnarray}
    \label{eq:gap1}
    \mathcal{M}_\sigma^2&=& \lambda \Bigl(3\phi_0^2-v^2
      + 3\hbar\, \mathcal{F}_\sigma 
      + (N-1) \hbar\, \mathcal{F}_\pi \Bigr)\\
    \label{eq:gap2}
    \mathcal{M}_\pi^2&=& \lambda \Bigl(\phi_0^2-v^2
      + \hbar\, \mathcal{F}_\sigma
      + (N+1) \hbar\, \mathcal{F}_\pi \Bigr)\ ,
  \end{eqnarray}
where $\mathcal{F}_*$ is the fluctuation integral
\[ \mathcal{F}_*(t) = \int \ddk{*} |f_*(k,t)|^2
     \quad\textrm{with}\quad *=\sigma,\pi
\] 
which equals the usual tadpole integral at $t=0$ 
(cf. section 1.4).
The equation for the mode functions is
\begin{equation}
  \label{eq:Modengleichung}
  \ddot{f}_*(k,t) + \biggl[ \vvec{k}^2
  + \mathcal{M}^2_{*}(t)  \biggr] f_*(k,t) = 0\ .
\end{equation}
The fact that the mode equation is coupled to the gap equations
\Gl{eq:gap1} and \Gl{eq:gap2} has an important influence on the
dynamics.
When a time-dependent mass square $\mathcal{M}^2(t)$ acquires a
negative value, eq.~\Gl{eq:Modengleichung} will imply an exponential
growth of the modes which reacts back on the mass squares via the 
gap equations by driving them back to positive values.
In the one-loop approximation (with no gap equations) the system
shows unphysical behavior \cite{Baacke:1997rcs} because the modes
never stop growing exponentially.

\subsection{Initial conditions}
\label{sec:initial}
At the beginning of the nonequilibrium evolution we fix the 
classical background field to a certain value
$\phi(0)=\phi_0$. The mode functions
are those of free fields:
$f_i(k,0) = 1$, $\dot f_i(k,0) = -i \omega_i$, 
and the mass parameters $\mathcal{M}_\sigma$ and $\mathcal{M}_\pi$
are solutions of the gap equations \Gl{eq:gap1} and \Gl{eq:gap2}
at $t=0$.

\section{Results and discussion}
\subsection{Time evolutions}
\begin{figure}[htbp]
  \centering
    \psfragscanon
    \psfrag{f}{\scriptsize $\phi(t)$}
    \psfrag{t}{\scriptsize $t$}
    \includegraphics[scale=0.2,angle=270]{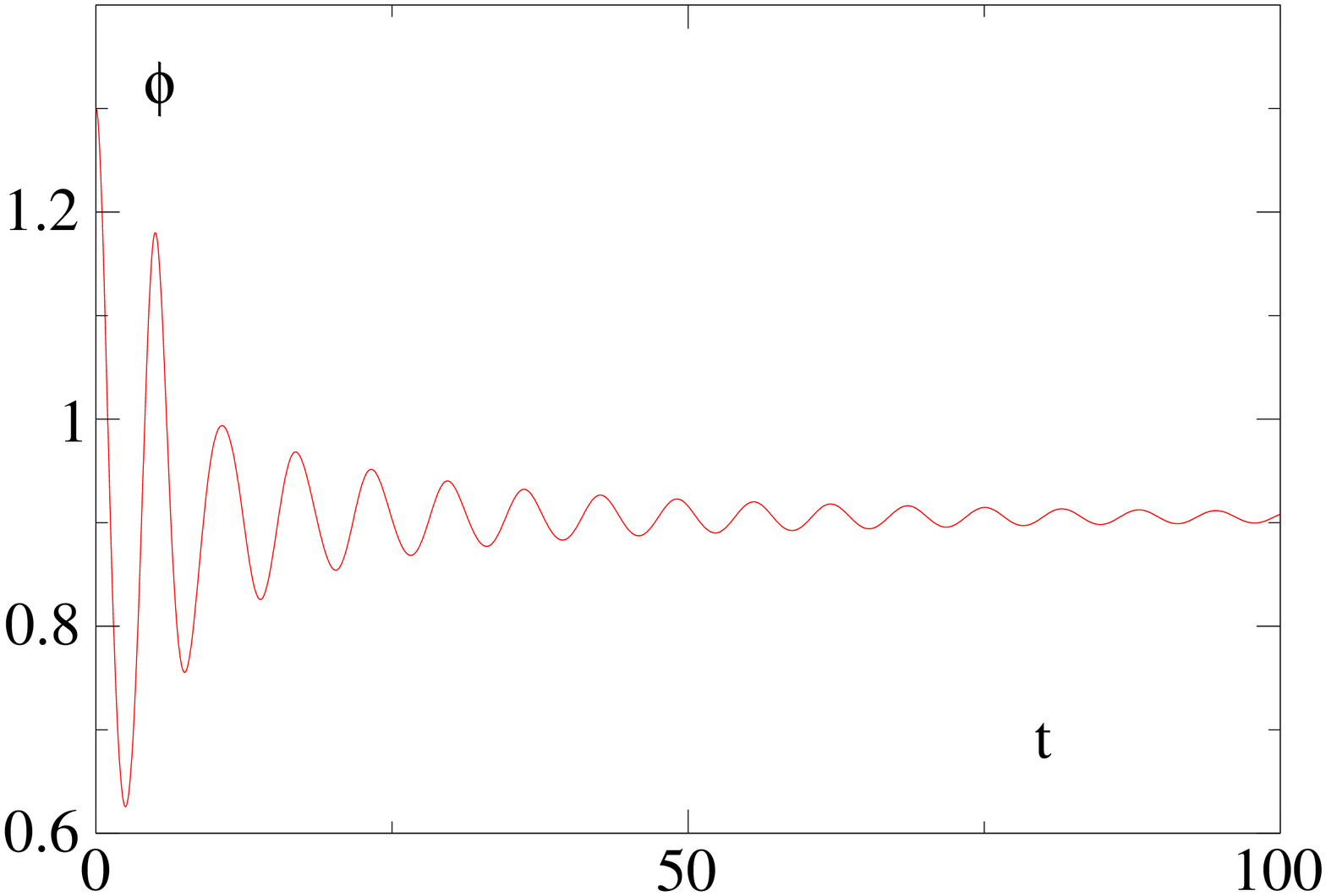}
    \includegraphics[scale=0.2,angle=270]{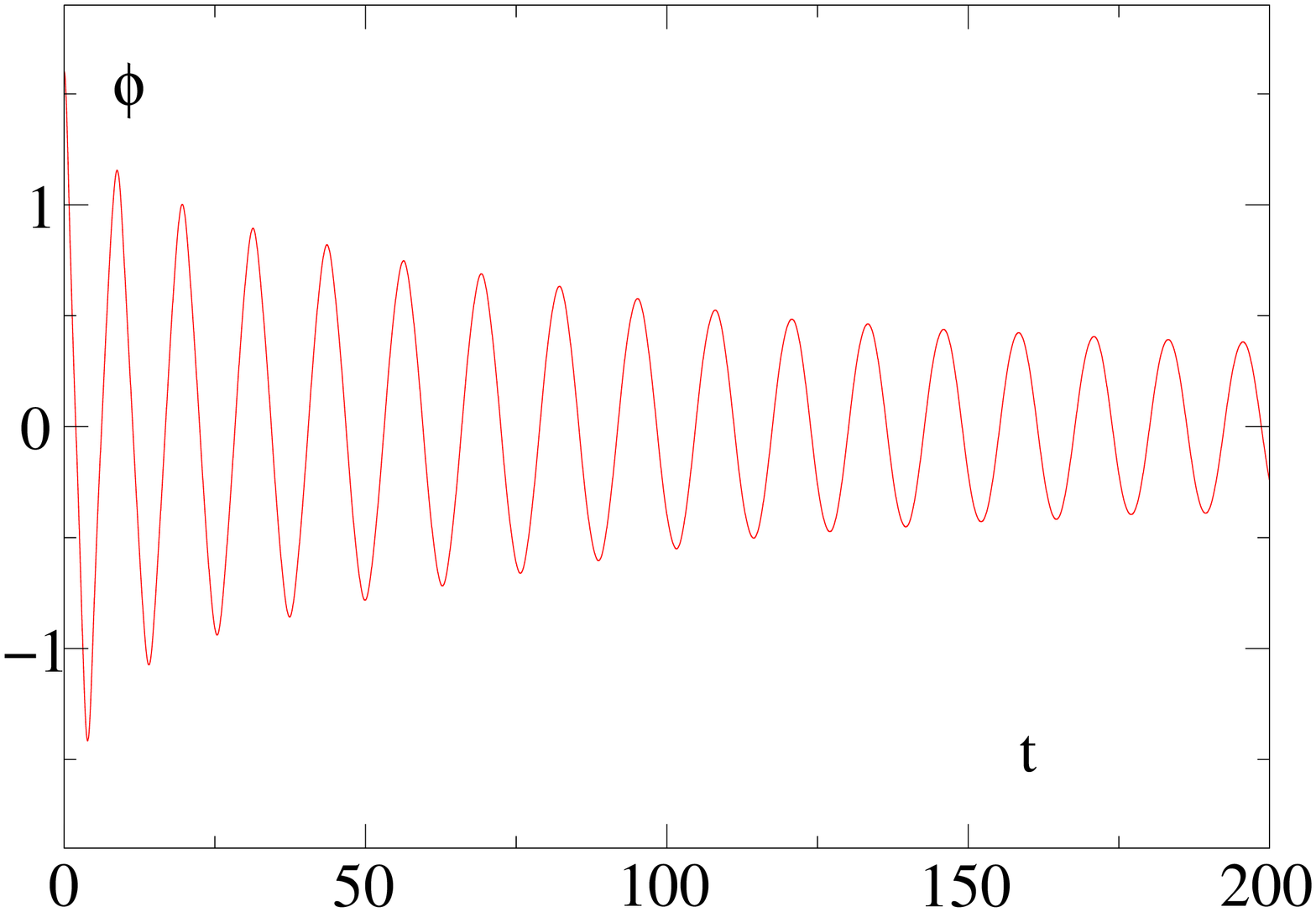}
    \caption{Time evolutions of the background field
      for $\phi_0=1.3v$, $\phi_0=1.6v$. 
      Amplitudes are in units of $v$ and
      time is measured in units of $(\sqrt{\lambda}v)^{-1}$.}
    \label{fig:phi(t)}
\end{figure}

Here we will present the results for $N=4$, $\lambda=1$ and for the 
renormalization scale set equal to the tree level sigma mass
$\mu^2=2\lambda v^2$.
We have only
considered initial values $\phi_0> v$ for the background 
field because
for smaller values the initial value of the parameter $\mathcal{M}_\pi(0)$
is imaginary. This means that the region $-v < \phi < v$ can only be 
explored dynamically.
We display time evolutions of the background field for two different
initial conditions in Fig.~\ref{fig:phi(t)}

It can be seen that there are two phases depending on the value of $\phi_0$:
a symmetric phase when the field $\phi(t)$ is oscillating about 
zero, and a phase of broken symmetry when $\phi(t)$ is oscillating
about a finite value. The ``critical'' value of $\phi_0$ seems to
be close to the classically expected value $\sqrt{2\lambda} v$
where the total energy is equal to the height of the barrier.
  
\subsection{Phase structure}
In order to analyze the phase structure of the model in the
\textnormal{Hartree} approximation we define $\phi_\infty$ as
a time averaged amplitude at late times, i.e. the value about which
the field oscillates. $\phi_\infty$ plays the role of an order parameter
whose dependence on the 
total energy of the system is investigated.
The total energy of the system is given here by the initial value $\phi_0$ 
which is analogous to the temperature in equilibrium field theory.
\begin{figure}[htbp]
  \centering
  \psfragscanon
  \psfrag{f}{\footnotesize $\phi_\infty$}
  \psfrag{f_0}{\footnotesize $\phi_0$}
  \psfrag{inf}{}
  \includegraphics[scale=0.2,angle=270]{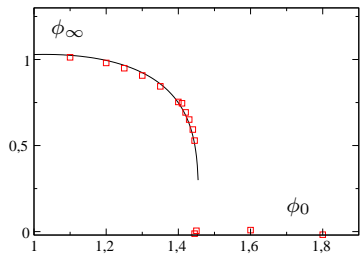}
  \caption{The late-time amplitude as function of the initial
  amplitude for $N=4$.}

  \psfragscanoff
  \label{fig:phase}
\end{figure}

Fig.~\ref{fig:phase} clearly shows a discontinuous jump of $\phi_\infty$
at $\phi_0\simeq \sqrt{2}v$ --- a typical sign of a first order
phase transition as found in equilibrium (see e.g. 
Refs.~\cite{Verschelde:2001onft,Nemoto:1999onft}).

\subsection{A dynamical effective potential}
\begin{figure}[htbp]
  \centering
  \psfragscanon
  \psfrag{f}{\scriptsize $\phi$}
  \psfrag{V}{\scriptsize $V_\mathrm{pot}$}
  \includegraphics[scale=0.20,angle=270]{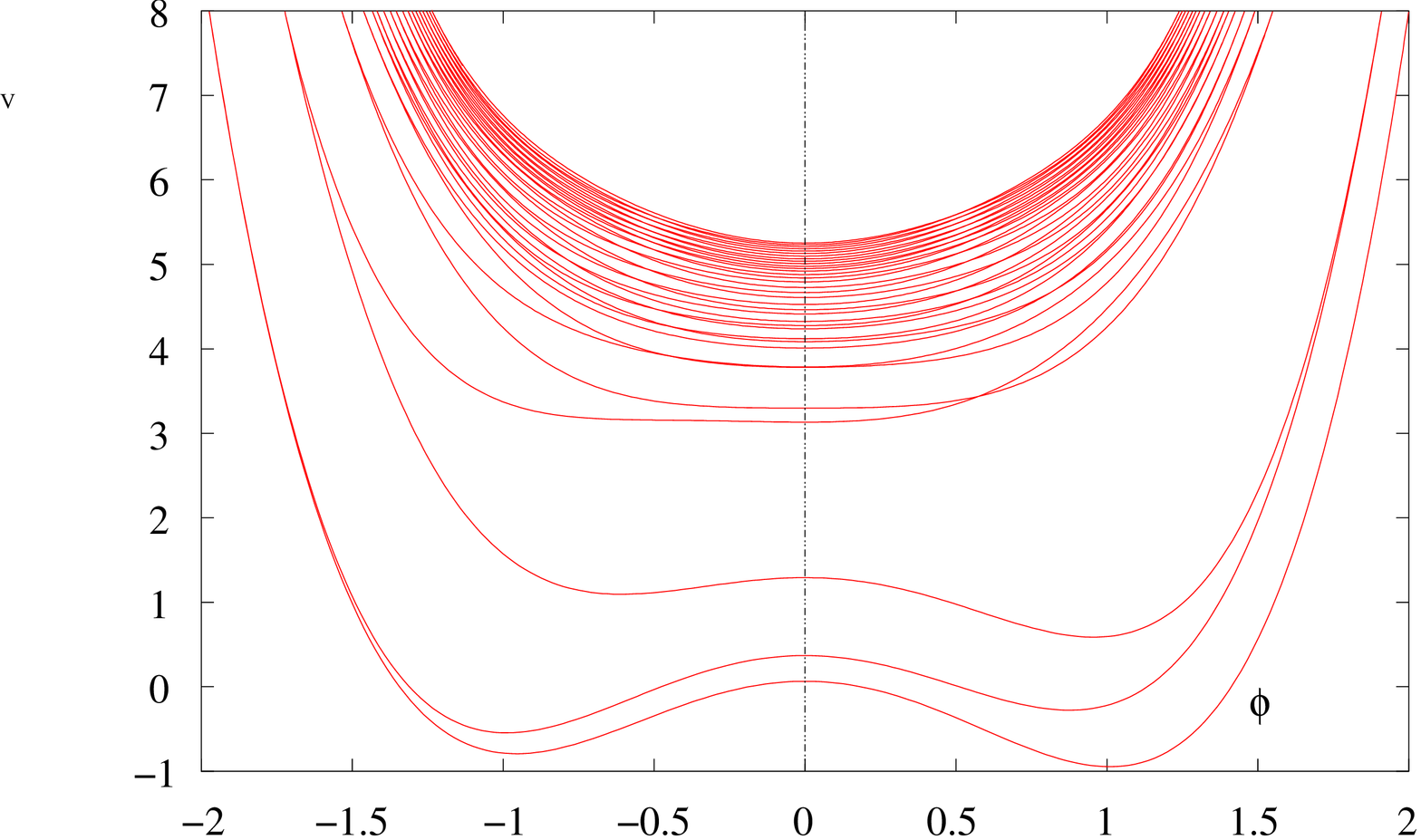}
  \hspace*{.5cm}
  \includegraphics[scale=0.20,angle=270]{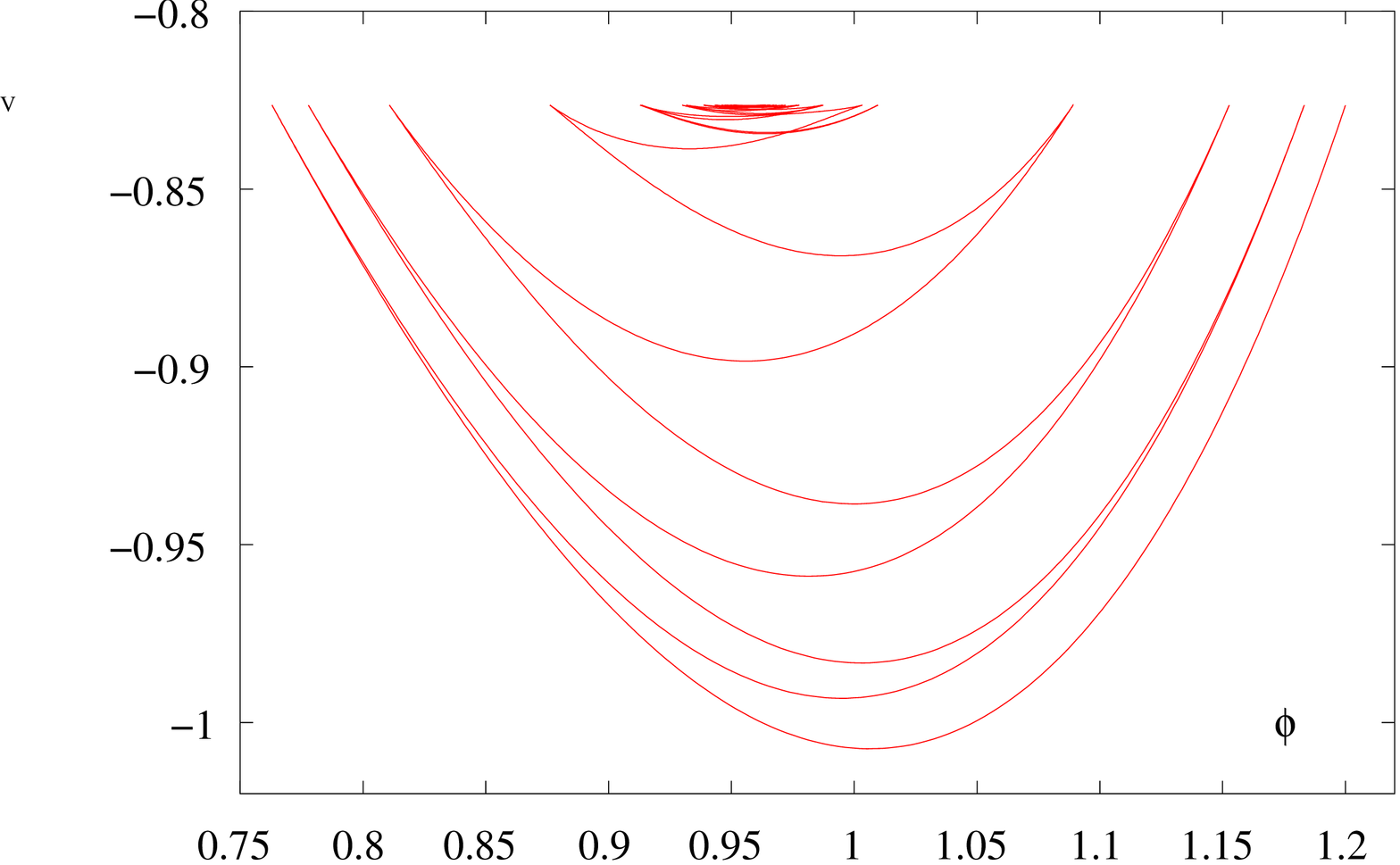}
  \psfragscanoff

  \caption{Evolution of the potential energy \Gl{eq:potenergie}
    for $\phi_0=2v$ (unbroken
    phase)  and $\phi_0=1.2v$ (broken phase).}
  \label{fig:V(phi)}
\end{figure}

This nonequilibrium system shows a phase structure 
which is comparable to a system in thermal 
equilibrium, so it would be nice if there was another correspondence.
We define a dynamical, i.e. time-dependent, 
potential which can be compared to the finite temperature potential 
in equilibrium
\begin{equation}
  \label{eq:potenergie}
  V_\mathrm {pot}(t) = \mathcal{E} - \halb \dot\phi^2(t)\ .  
\end{equation}
This potential can only be measured within the oscillation range of
the background field $\phi(t)$. For two different initial conditions
it is shown in Fig.~\ref{fig:V(phi)}.
In the broken symmetry phase the minimum at $\phi=v$
moves to smaller values but eventually remains different from zero, so the
system settles at a finite expectation value.
In the symmetric phase the two minima entirely disappear after a few
oscillations and a new (symmetric) minimum at $\phi=0$ appears.

\section{Conclusions and summary}
The analysis of the nonequilibrium dynamics of the $O(N)$ model in
the \textnormal{Hartree} approximation allowed us to study new features
of the system which are not accessible in the one-loop or infinite
component ($N\to\infty$) approximations.
Though thermalization is only expected at approximations beyond the 
\textnormal{Hartree} level, i.e., when including nonlocal
corrections, the nonequilibrium system at late times 
shows striking similarities to a system in thermal equilibrium.
One can define an order parameter which is dependent on the 
total energy of the system, given by the initial conditions.
Analyzing the dependence of the order parameter on the initial conditions
one finds a first order phase transition.

\end{document}